\documentclass[12pt]{iopart}
\newtheorem{twr}{Theorem}
\usepackage{color}
\usepackage{graphicx}
\usepackage{epsf}

\def\be{\begin{equation}}
\def\ee{\end{equation}}
\def\bea{\begin{eqnarray}}
\def\eea{\end{eqnarray}}
\def\bi{\begin{itemize}}
\def\ei{\end{itemize}}
\def\bin{\begin{enumerate}}
\def\ein{\end{enumerate}}

\def\ra{\rangle}
\begin{document}
\title[Inert-states of spin-5 and spin-6 Bose-Einstein condensates]
{Inert-states of spin-5 and spin-6 Bose-Einstein condensates}

\author{Marcin Fizia and Krzysztof Sacha}

\address{Instytut Fizyki imienia Mariana Smoluchowskiego \\ and
Mark Kac Complex Systems Research Center, \\
Uniwersytet Jagiello\'nski, \\ ulica Reymonta 4, PL-30-059 Krak\'ow, Poland}

\begin{abstract}
In this paper we consider spinor Bose-Einstein condensates with spin $f=5$
and $f=6$ in the presence and absence of external magnetic field at the mean field level. We calculate all of so-called inert-states of these systems. Inert-states are very unique class of stationary states because they remain stationary while Hamiltonian parameters change. Their existence comes from Michel's theorem. For illustration of symmetry properties of the inert-states 
we use method that allows classification of the systems as a polyhedron with $2f$ vertices proposed by R. Barnett {\it et al.}, Phys. Rev. Lett. {\bf 97}, 180412 (2006).
\end{abstract}
\pacs{03.75.Mn,03.75.Hh}
\maketitle
\section{Introduction}
Properties of Bose-Einstein condensates have been being found vastly since 1995 when first condensate was performed experimentally\cite{wytw1, wytw2, wytw3, peth, art}. After optical dipole traps were developed it was possible to create not only scalar but also spinor condensates. Computing properties of spinor condensates is more complicated than these of scalar one \cite{peth, art, ueda}. In scalar condensates only one $s$-wave channel of interaction is possible since only one kind of bosons is present. On the other hand in spinor condensate of spin-$f$ atoms there are $2f+1$ internal states of atom. Therefore number of interaction channels increases. Because of this complexity finding stationary states for spinor condensates is not trivial even within the mean field approximation. 

In this paper we will first briefly show how to construct Hamiltonian for spinor condensates and how to perform calculations within mean field theory for that system \cite{ueda}. After that we shall present Michel's theories and point out how it could help us finding stationary states \cite{Micheltw}. The Michel's theorem ensures us that there have to exist so-called inert-states. Their name comes from a fact that they remain stationary for every values of Hamiltonian's parameters (such as external magnetic field, density etc.). Finally we will show how to use Michel's theorem and how to perform calculations of stationary states of spinor condensates \cite{ueda,yip, makela}. All inert states for spin $f\le  4$ condensates and some inert states for higher spins have been presented in the literature \cite{ueda,yip, makela}. We will present all inert states for spin-5 and spin-6 condensates with and without external magnetic field. For illustration of symmetry properties of the states we use spherical harmonics expansion and the method proposed by Barnett 
{\it et. al} \cite{deml,deml1,com}.

\section{Hamiltonian and mean field theory}
That part is very well described in \cite{ueda} so we shall briefly point way for constructing Hamiltonian and its most important assumptions. In spinor condensates atoms possess internal degrees of freedom characterized by a state of hyperfine structure. We should remember that usually in spinor Bose-Einstein condensates description by $f$ we understand total angular momentum of all electrons and nuclei of atom which we will just call spin.
So the wave function of spinor condensate in the mean field description should be written as:
\begin{equation}
\Psi(\vec{r}_{1},...,\vec{r}_{N})=\prod_{i=1}^N 
\left[
\begin{array}{c}
\Psi_f(\vec r_i) \\
\vdots \\
\Psi_{-f}(\vec r_i)
\end{array}
\right]\;,
 \label{eq:kondspinfunkfalo}
\end{equation}
where $\Psi_{m}(\vec r_i)$ is wave function component describing particles with projection $m$ of spin on chosen quantization axis.

Let us denote $\psi_{m}^{\dag}(\vec{r})$ as a bosonic field operator which creates a boson at point $\vec{r}$ with spin projection equal $m$. 
In the absence of external trapping potential but in the presence of external magnetic field, energy of this system can be decomposed into kinetic energy, energy of Zeeman effects (linear and quadratic) and interaction energy. So we can write:
\begin{equation}
 \hat{H}=\hat{H_0}+\hat{V}\;,
\label{eq:h}
\end{equation}
where by $\hat{H}_{0}$ we understand part of Hamiltonian without interaction between particles. Assuming the direction of external magnetic field $\vec B$ 
along the $\vec{z}$-axis it can be decomposed as follows:
\begin{equation}
\hat{H}_0=\int d^3 \vec{r} \sum_{m_1,m_2=-f}^{f}\hat{\psi}_{m_1}^\dag (\vec{r})\left[-\frac{\hbar^{2} \bigtriangledown^2}{2\tilde M}-p(f_z)_{m_1m_2}+q(f_z^2)_{m_1m_2}\right]\hat{\psi}_{m_2}(\vec{r})\;,
 \label{eq:hamigoly1}
\end{equation}
where $\tilde M$ is mass of atoms, $p\sim B$ factor describes the linear Zeeman effect, $q\sim B^2$ factor describes the quadratic Zeeman effect and $(f_z)_{m_1m_2}$ is spin matrix (in the chosen coordinates system it is diagonal $(f_z)_{m_1m_2}=m_1\delta_{m_1,m_2}$). 

In dilute and ultra-cold atomic gases two-body $s$-wave interaction channels are important only. 
From wave function's symmetry considerations we can find out that the only possible channels for identical bosons correspond to even values
of total spin $F$ of two interacting particles.
So the interaction part of the Hamiltonian can be decomposed in the following way 
\begin{equation}
\hat{V}=\sum_{F=0,2,...,2f}\hat{V}^{(F)}\;.
 \label{eq:oddz1}
\end{equation}
We can construct (using Clebsh-Gordon coefficients) operators which annihilate pairs of bosons with total angular momentum $F$ and its projection $M$ on the $\vec{z}$-axis
\begin{equation}
\hat{A}_{FM}(\vec{r},\vec{r}\;')=\sum_{m_1,m_2=-f}^f \langle F,M|f,m_1;f,m_2\rangle\hat{\psi}_{m_1}(\vec{r})\hat{\psi}_{m_2}(\vec{r}\;')\;.
 \label{eq:anihpary}
\end{equation}
Now we can see that operators $\hat{V}^{(F)}$ can be written as:
\begin{equation}
\hat{V}^{(F)}=\frac{1}{2}\int d^3 r \int d^3{r}' \nu^{(F)}(\vec{r},\vec{r}\;')\sum_{M=-F}^F \hat{A}^\dag_{FM}(\vec{r},\vec{r}\;')\hat{A}_{FM}(\vec{r},\vec{r}\;')\;.
 \label{eq:oddzF}
\end{equation}
Function $\nu^{(F)}(\vec{r},\vec{r}\;')$ describes spatial dependence of interactions between two particles in the channel $F$. This function using $s$-wave scattering length $a_F$ can be assumed to be:
\begin{equation}
\nu^{(F)}(\vec{r},\vec{r}\;')=g_F \delta^{(3)}(\vec{r}-\vec{r}\;')\;,
 \label{eq:nu}
\end{equation}
where $g_F=\frac{4\pi \hbar^2}{M}a_F$.

We can also use two relations to simplify our final form of the 
Hamiltonian. Namely:
\begin{itemize}
 \item Hilbert space completeness relation:
\begin{equation}
\sum_{F=0,2,...,2f}\sum_{M=-F}^F \hat{A}_{FM}^\dag(\vec{r})\hat{A}_{FM}(\vec{r})=:\hat{n}(\vec{r})\hat{n}(\vec{r}):\;,
 \label{eq:zupel}
\end{equation}
where $:\;:$ stands for normal ordering of the creation and annihilation operators,
\item relation coming from composition of angular momentum
\begin{equation}
:\hat{\vec{F}}(\vec{r})\cdotp\hat{\vec{F}}(\vec{r}):=\sum_{F=0,...,2f}\sum_{M=-F}^F\left[\frac{1}{2}F(F+1)-f(f+1)\right]\hat{A}_{FM}^\dag(\vec{r})\hat{A}_{FM}(\vec{r})\;.
 \label{eq:skladanie}
\end{equation}
\end{itemize}

In the particle number conserving version of the mean field theory we have to calculate mean value of the Hamiltonian assuming that $N$ bosons are in the same state, see Eq.~(\ref{eq:kondspinfunkfalo}). We assume also that the spinor and spatial parts of the wave-function factorize, i.e. $|\Psi(\vec r)\ra=\phi_0(\vec r)|\zeta\ra$. In the absence of an external trapping potential the ground state mode 
$\phi_0=\frac{1}{\sqrt{\cal V}}$, where $\cal V$ is the volume of the system,
and the energy reduces to the expression
\begin{equation}
 E(\zeta)=\langle\zeta|\hat{H}|\zeta\rangle\;,
\label{eq:wartoscsrednia}
\end{equation}
where $|\zeta\rangle=\frac{1}{\sqrt{N!}}\left(\sum_{m=-f}^f \zeta_{m}\hat{a}^\dag_{m;0}\right)^N|0\rangle\;$.
Vector $\zeta=(\zeta_F,...,\zeta_{-F})^T$ will be called an order parameter. This parameter completely describes state of spinor condensate. 
Neglecting terms proportional to $1/N$ final result for the energy per particle reads \cite{ueda}
\bea
\varepsilon(\zeta)=\frac{E(\zeta)}{N}&=&\sum_{m=-f}^f \left[
|\zeta_m|^2(-pm+qm^2)+\frac{1}{2}\rho c_0+\frac{1}{2}\rho c_1 |\vec{F}|^2 \right.
\cr &&
\left. +\sum_{F=0,...,2f-4}\frac{d_F \rho }{2}\sum_{M=-F}^F A^{\ast}_{FM}A_{FM}\right]\;,
 \label{eq:ennaczastke}
\eea
where $\rho$ is the particle density. 
Parameters $c_0$, $c_1$, $d_i$ are linear combinations of coefficients $g_i$ and are presented in Appendix 1. Functions used in (\ref{eq:ennaczastke}) have the following forms
\begin{equation}
 F_{\mu}=\sum_{m_1,m_2=-f}^f \zeta_{m_1}^\ast(f_\mu)_{m_1m_2} \zeta_{m_2}\;,
\label{eq:spinor}
\end{equation}

\begin{equation}
 A_{FM}=\sum_{m_1,m_2=-f}^f\langle F,M|f,m_1;f,m_2\rangle\zeta_{m_1}\zeta_{m_2}\;.
\label{eq:afmprim}
\end{equation}

In order to calculate all stationary states of the energy (\ref{eq:ennaczastke}) one has to derive the corresponding Euler-Lagrange equations and solve them which usually has to be done numerically. However, a certain class of stationary states can be found analytically by consideration of symmetries of the system. These are the so-called inert-states which we present in the following sections.

\section{Michel's theorem \cite{Micheltw}}

Michel's theorem allows us to find special stationary states. They are called inert-states because they are stationary for any values of Hamiltonian's parameters (in our case the parameters are: particle density,  scattering lengths and external magnetic field). The stationarity of these states is guaranteed by symmetries of a system. 

When external magnetic field is present the symmetry group of the energy function (\ref{eq:ennaczastke}) is a product group $G_{B\neq0}=U(1)_{F_z} \times U(1)$. Part $U(1)_{F_z}$ is rotation about the field direction (i.e. $\vec{z}$-axis) and part $U(1)$ is just global phase changing symmetry. If an external magnetic field is not present the symmetry group is $G_{B=0}=SO(3) \times U(1)$. Meaning of the part $U(1)$ is the same, i.e. it is the global phase changing symmetry, whereas $SO(3)$ is group of rotations in 3-dimensional space.

For our further considerations we should call symmetry group by $G$ and by $\varepsilon(\zeta)$ function defined in a finite-dimensional smooth domain $\cal M$ which has this symmetry. By orbit $G(\zeta)$ of point $\zeta$ we understand set of points being results of action of all group elements on point $\zeta$
\[G(\zeta)=\{g \zeta| g \in G\}
\subset {\cal M}
\;.\]
Isotropy group of point $\zeta$ is a set of group elements which act on $\zeta$ as an identity element
\[G_\zeta=\{g\in G|g \zeta =\zeta\}\subset G\;.\]
We say that any two isotropy groups are conjugate if
\[\exists g\in G: G_\zeta=g \cdot G_{\zeta'}\cdot  g^{-1}\;.\]
Stratum is union of all orbits of points whose isotropy groups are conjugate
\[S(\zeta)=\bigcup_{\zeta':G_{\zeta}=g G_{\zeta'}g^{-1}} G(\zeta')
\subset {\cal M}
\;.\]
We say that an orbit is isolated in its stratum if there exists neighborhood of this orbit $U\supset G(\zeta)$ so that its intersection with stratum gives that orbit
\[G(\zeta)=U \cap S(\zeta)\;.\]
Michel proved following theorem.
\begin{twr}[Michel]
 Every function defined in a finite-dimensional smooth domain $\cal M$ and having real numbers as co-domain which is invariant under elements of compact Lie group $G$ i.e.:
\[\forall \varepsilon: {\cal M}\rightarrow \Re: \forall g\in G \wedge \forall \zeta\in {\cal M}: \varepsilon(\zeta)=\varepsilon(g\zeta)\;,\] 
have common orbits of extrema which are orbits being isolated in their stratum.
\end{twr}

From this theorem we know that for our spinor condensate system we are able to find some stationary points $\zeta$ just by finding isolated orbits. The crucial point of the Michel's theorem proof is to show that gradient of function $\varepsilon(\zeta)$ is a vector which is tangential to the stratum. We know from definition of orbit that gradient of function is orthogonal to the orbit. And if any orbit is isolated in its stratum we know that gradient has to be null on it so this orbit has to be extremal (minimal, maximal or saddle) point. 

Barnett {\it et. al.} \cite{deml} showed how to translate symmetries of spin-$f$ states $|\zeta\rangle$ into symmetries of a polyhedron with $2f$ vertices. We should not redo it here but point out the most important steps. We have to find a set of maximally polarized states, i.e. $\hat{\vec{f}}|\chi\ra=f|\chi\rangle$, which are orthogonal to $|\zeta\ra$,
\begin{equation}
 \langle\zeta|\chi\rangle=0\;.
\label{eq:deml1}
\end{equation}
The states $|\chi\ra$ can be parameterized as $\chi=e^{i \phi}\tan{\frac{\theta}{2}}$ where $\phi$ and $\theta$ denote position of  vertices of a polyhedron on the unit sphere.
For our considerations the most important corollary is that maximal number of found vertices is equal to $2f$ because Eq.~(\ref{eq:deml1}) is actually a polynomial equation of the $2f$ degree. Therefore when we are looking for inert-states of the system we may consider only these point group whose corresponding regular polyhedron has no more than $2f$ vortices.

Now we should present way of calculating and finding inert-states \cite{ueda, yip, makela}. In order to do that we should go through all subgroups of a global symmetry group of a system. For each subgroup we should find states which are invariant under this subgroup. In that case this subgroup is their isotropy group. 

First we will show how to calculate inert-states for subgroups with only one generator. In our case such generators have form $\hat{K}=e^{i \hat f_{\Omega}\phi}$ (i.e. they are rotation operator about $\vec{\Omega}$-axis by an angle $\phi$). It is worth pointing out that if $\hat{K}$ belongs to the  symmetry group of the system so is $e^{i\lambda}\hat{K}$ because in our case the symmetry group of the system is a direct product of a rotation group and the global phase changing symmetry group. All eigenvalues of $\hat{K}$ can be written as $e^{i\gamma}$. If we find all eigenvectors of a generator $\hat{K}$ we can identify some of them to be inert-states. 
Through eigensubspaces of $\hat{K}$ we should choose these which are non-degenerate. That means that for each eigenvalue $e^{i\gamma}$ there exists only one eigenvector $\zeta_{\gamma}$. Therefore we know that group generated by $e^{-i\gamma}\hat{K}$ is isotropy group of this vector. Moreover we see that there is only one orbit whose isotropy group is this group. Thus, this orbit must be isolated in its stratum because stratum consists of this orbit only. Then, the Michel theorem tells us $\zeta_{\gamma}$ is an inert-state. 

If some subgroup has two or more generators, as in the case of point groups, we ought to find vectors which are common eigenvectors for all generators. If there is only one common eigenvector it is an inert-state of the system.
When on the other hand the common eigenspace is not one-dimensional 
we still can calculate stationary states which are not inert-states. To this end we have to find extrema of the energy of the system within the common eigensubspace.
Existence of such stationary states comes from another Michel's theorem \cite{ueda}. However, these states depend on the Hamiltonian parameters. We shall not consider them in this paper.

\section{Inert-states for spin-5 condensates with external magnetic field}

\begin{figure}[h]
 \centering
 \includegraphics[width=6.cm,keepaspectratio=true]{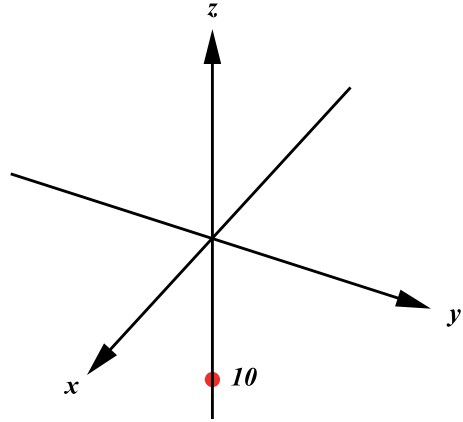}
 \caption{Visualization of the state $\zeta_{F5+}$, Eq.~(\ref{eq:5f5+}), with Barnett {\it et al.} method \cite{deml}. Red circles indicate positions of $2f=10$ vertices of a polyhedron. In the presented case all vertices are situated on the same point (that is indicated by the number 10 in the figure) and the polyhedron is reduced to a single point.}
 \label{Fig:5f5}
\end{figure}
\begin{figure}[h]
 \centering
 \includegraphics[width=6.cm,keepaspectratio=true]{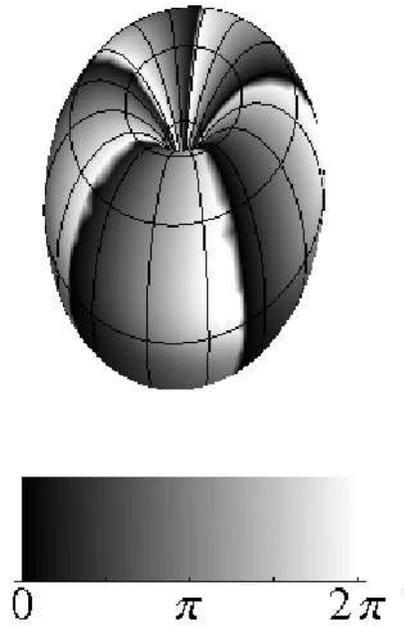}
 \caption{Visualization of the state $\zeta_{F5+}$, Eq.~(\ref{eq:5f5+}),  by means of spherical harmonics. Surface $|\Psi(\theta,\phi)|^2=1$, see Eq.~(\ref{eq:funkcjafalowaspinowa}), is plotted in the figure. Colours express phase of the state at a given point.}
 \label{Fig:5f5sph}
\end{figure}

Global symmetry group of spinor condensates with external magnetic field is direct product group $ G_{B\neq0}=U(1)_{F_z} \times U(1)$. General group element has form:
\begin{equation}
g(\phi,\gamma)=e^{i \phi}e^{-i \hat{f}_z \gamma}\;,
 \label{eq:elgr5bez}
\end{equation}
where $\phi$ is global phase change and $\gamma$ is angle of rotation about the $\vec{z}$-axis which coincides with the direction of the magnetic field. We have found 11 inert-states for this case. All of them are inert-states whose isotropy groups are continuous. Analysis of discrete subgroups does not reveal any additional inert-state. 

We present the inert-states with their isotropy group's generator and the corresponding energy. Isotropy groups are denoted by $U(1)_{F_z+\alpha(\phi)}$. Only for one state we have decided to illustrate its symmetry. One way is that proposed by Barnett {\it et. al.} \cite{deml}. Second one is by using spherical harmonic expansion. That is, every bosonic spin state can be expand in terms of spherical harmonics as follows
\begin{equation}
 \Psi(\theta, \phi)=\sum_{m=-f}^f \zeta_m Y_f^m(\theta, \phi)
\label{eq:funkcjafalowaspinowa}
\end{equation}
We can visualize states by painting surfaces $|\Psi(\theta,\phi)|^2=1$. Colour of surface should express phase of state at a given point \cite{deml, com}.
\begin{itemize}
 \item $U(1)_{F_z+5\phi}$

generator: $\{e^{5i\phi}e^{-i\phi \hat{f}_z}\}$

\begin{equation}
 \zeta_{F5+}=(1,0,0,0,0,0,0,0,0,0,0)^T
\label{eq:5f5+}
\end{equation}
\begin{equation}
 \varepsilon(\zeta_{F5+})=\frac{1}{2}\rho c_0+\frac{25}{2}\rho c_1+25 q-5p\;,
\label{eq:en5f5+}
\end{equation}
The state is visualized in Fig.~\ref{Fig:5f5} and Fig.~\ref{Fig:5f5sph}.

\item $U(1)_{F_z+4\phi}$

generator: $\{e^{4i\phi}e^{-i\phi \hat{f}_z}\}$

\begin{equation}
 \zeta_{F4+}=(0,1,0,0,0,0,0,0,0,0,0)^T
\label{eq:5f4+}
\end{equation}
\begin{equation}
 \varepsilon(\zeta_{F4+})=\frac{1}{2}\rho c_0+8\rho c_1+16q-4p\;,
\label{eq:en5f4+}
\end{equation}
\item $U(1)_{F_z+3\phi}$

generator: $\{e^{3i\phi}e^{-i\phi \hat{f}_z}\}$

\begin{equation}
 \zeta_{F3+}=(0,0,1,0,0,0,0,0,0,0,0)^T
\label{eq:5f3+}
\end{equation}
\begin{equation}
 \varepsilon(\zeta_{F3+})=\frac{1}{2}\rho c_0+\frac{9}{2}\rho c_1+\frac{14}{85}\rho d_6+9q-3p\;,
\label{eq:en5f3+}
\end{equation}
\item $U(1)_{F_z+2\phi}$

generator: $\{e^{2i\phi}e^{-i\phi \hat{f}_z}\}$

\begin{equation}
 \zeta_{F2+}=(0,0,0,1,0,0,0,0,0,0,0)^T
\label{eq:5f2+}
\end{equation}
\begin{equation}
 \varepsilon(\zeta_{F2+})=\frac{1}{2}\rho c_0+2\rho c_1+\frac{35}{286}\rho d_4+\frac{84}{935}\rho d_6 +4q-2p\;,
\label{eq:en5f2+}
\end{equation}
\item $U(1)_{F_z+\phi}$

generator: $\{e^{i\phi}e^{-i\phi \hat{f}_z}\}$

\begin{equation}
 \zeta_{F1+}=(0,0,0,0,1,0,0,0,0,0,0)^T
\label{eq:5f1+}
\end{equation}
\begin{equation}
 \varepsilon(\zeta_{F1+})=\frac{1}{2}\rho c_0+\frac{1}{2}\rho c_1+\frac{25}{286}\rho d_2+\frac{10}{143}\rho d_4+\frac{14}{187}\rho d_6+q-p\;,
\label{eq:en5f1+}
\end{equation}
\item $U(1)_{F_z}$

generator: $\{e^{-i\phi \hat{f}_z}\}$

\begin{equation}
 \zeta_{P}=(0,0,0,0,0,1,0,0,0,0,0)^T
\label{eq:5p}
\end{equation}
\begin{equation}
 \varepsilon(\zeta_{P})=\frac{1}{2}\rho c_0+\frac{1}{22}\rho d_0+\frac{25}{429}\rho d_2+\frac{9}{143}\rho d_4+\frac{40}{561}\rho d_6\;,
\label{eq:en5p}
\end{equation}
\item $U(1)_{F_z-\phi}$

generator: $\{e^{-i\phi}e^{-i\phi \hat{f}_z}\}$

\begin{equation}
 \zeta_{F1-}=(0,0,0,0,0,0,1,0,0,0,0)^T
\label{eq:5f1-}
\end{equation}
\begin{equation}
 \varepsilon(\zeta_{F1-})=\frac{1}{2}\rho c_0+\frac{1}{2}\rho c_1+\frac{25}{286}\rho d_2+\frac{10}{143}\rho d_4+\frac{14}{187}\rho d_6 +q+p\;,
\label{eq:en5f1-}
\end{equation}
\item $U(1)_{F_z-2\phi}$

generator: $\{e^{-2i\phi}e^{-i\phi \hat{f}_z}\}$

\begin{equation}
 \zeta_{F2-}=(0,0,0,0,0,0,0,1,0,0,0)^T
\label{eq:5f2-}
\end{equation}
\begin{equation}
 \varepsilon(\zeta_{F2-})=\frac{1}{2}\rho c_0+2\rho c_1+\frac{35}{286}\rho d_4+\frac{84}{935}\rho d_6+4q+2p\;,
\label{eq:en5f2-}
\end{equation}
\item $U(1)_{F_z-3\phi}$

generator: $\{e^{-3i\phi}e^{-i\phi \hat{f}_z}\}$

\begin{equation}
 \zeta_{F3-}=(0,0,0,0,0,0,0,0,1,0,0)^T
\label{eq:5f3-}
\end{equation}
\begin{equation}
 \varepsilon(\zeta_{F3-})=\frac{1}{2}\rho c_0+\frac{9}{2}\rho c_1+\frac{14}{85}\rho d_6+9q+3p\;,
\label{eq:en5f3-}
\end{equation}
\item $U(1)_{F_z-4\phi}$

generator: $\{e^{-4i\phi}e^{-i\phi \hat{f}_z}\}$

\begin{equation}
 \zeta_{F4-}=(0,0,0,0,0,0,0,0,0,1,0)^T
\label{eq:5f4-}
\end{equation}
\begin{equation}
 \varepsilon(\zeta_{F4-})=\frac{1}{2}\rho c_0+8\rho c_1+16q+4p\;,
\label{eq:en5f4-}
\end{equation}
\item $U(1)_{F_z-5\phi}$

generator: $\{e^{-5i\phi}e^{-i\phi \hat{f}_z}\}$

\begin{equation}
 \zeta_{F5-}=(0,0,0,0,0,0,0,0,0,0,1)^T
\label{eq:5f5-}
\end{equation}
\begin{equation}
 \varepsilon(\zeta_{F5-})=\frac{1}{2}\rho c_0+\frac{25}{2}\rho c_2 +25q+5p\;.
\label{eq:en5f5-}
\end{equation}
\end{itemize}

\section{Inert-states for spin-5 condensates without external magnetic field}

In this case global symmetry group is $G_{B=0}=SO(3)\times U(1)$. General group element can be written as:
\begin{equation}
 g(\alpha,\beta,\gamma,\phi)=e^{i\phi}e^{-i\alpha \hat f_z}e^{-i\beta \hat f_y}e^{-i\gamma \hat f_z}\;,
\label{eq:elgr5z}
\end{equation}
where $\phi$ is global phase change and $\alpha$, $\beta$, $\gamma$ are Euler angles. 
This group has the same continuous subgroups as the global symmetry group considered in the previous section. So we expect that the corresponding inert-states are the same. It easy to show that in the case without magnetic field energies of some of these inert-states are equal. That is, for $p=q=0$ we get $\varepsilon(\zeta_{F5+})=\varepsilon(\zeta_{F5-})$, $\varepsilon(\zeta_{F4+})=\varepsilon(\zeta_{F4-})$, $\varepsilon(\zeta_{F3+})=\varepsilon(\zeta_{F3-})$, $\varepsilon(\zeta_{F2+})=\varepsilon(\zeta_{F2-})$ and $\varepsilon(\zeta_{F1+})=\varepsilon(\zeta_{F1-})$. We can show that 
members of each pair can be transformed to each others by applying some  elements of the group $G_{B=0}$. So they belong to the same orbit and therefore they are actually the same states. We  have decided not to show these states one more time but present how to change one vector into another by action with group elements:
\begin{equation}
 g(0,\pi,0,0)\zeta_{F5+}=\zeta_{F5-}
\label{eq:5+=5-z}
\end{equation}
\begin{equation}
 g(0,\pi,0,\pi)\zeta_{F4+}=\zeta_{F4-}
\label{eq:4+=4-z}
\end{equation}
\begin{equation}
 g(0,\pi,0,0)\zeta_{F3+}=\zeta_{F3-}
\label{eq:3+=3-}
\end{equation}
\begin{equation}
 g(0,\pi,0,\pi)\zeta_{F2+}=\zeta_{F2-}
\label{eq:2+=2-z}
\end{equation}
\begin{equation}
 g(0,\pi,0,0)\zeta_{F1+}=\zeta_{F1-}\;.
\label{eq:1+=1-z}
\end{equation}

Now we should consider all discrete point groups. 
Let us begin with the dihedral group $D_{n}$. Generators of this group are $\hat{C}_{n,z}$ (rotation about $\vec{z}$-axis by angle $2 \pi/n$) and $\hat{C}_{2,x}$ (rotation about $\vec{x}$-axis by angle $\pi$ with simultaneous phase change by $\pi$). We would like to stress that in our consideration these generators will be supplemented with additional phase factors. This is necessary in order to obtain proper isotropy groups and we are allowed to do that because the symmetry group of our system contains global phase changing freedom.
It is worth mentioning that the dihedral group with infinite $n$ is also an isotropy group of the previously found state $\zeta_P$, Eq.~(\ref{eq:5p}). In the limit $n\rightarrow \infty$ the dihedral group becomes continuous. 
Other inert-states for the dihedral group are presented here:
\begin{itemize}
 \item $D_{10}$

generators: $\{e^{i\pi}\hat{\mathbf{C}}_{2,x},e^{i\pi}\hat{\mathbf{C}}_{10,z}\}$

\begin{equation}
 \zeta_{D10}=\frac{1}{\sqrt{2}}(1,0,0,0,0,0,0,0,0,0,-1)^T\;.
\label{eq:5d10}
\end{equation}
We can show that  $\hat{\mathbf{C}}_{2,x}\zeta_{D10}=e^{-i\pi}\zeta_{D10}$
and $\hat{\mathbf{C}}_{10,z}\zeta_{D10}=e^{-i\pi}\zeta_{D10}$. Thus, in order to obtain the isotropy group we have to supplement the original dihedral group generators with the phase factor $e^{i\pi}$.

\begin{equation}
 \varepsilon(\zeta_{D10})=\frac{1}{2}\rho c_0+\frac{1}{22}\rho d_0+\frac{75}{572}\rho d_2+\frac{9}{143}\rho d_4+\frac{15}{1496}\rho d_6\;.
\label{eq:en5d10}
\end{equation}

 \item $D_{8}$

generators: $\{e^{i\pi}\hat{\mathbf{C}}_{2,x},e^{i\pi}\hat{\mathbf{C}}_{8,z}\}$

\begin{equation}
 \zeta_{D10}=\frac{1}{\sqrt{2}}(0,1,0,0,0,0,0,0,0,-1,0)^T
\label{eq:5d8}
\end{equation}
\begin{equation}
 \varepsilon(\zeta_{D10})=\frac{1}{2}\rho c_0+\frac{1}{22}\rho d_0+\frac{3}{143}\rho d_2+\frac{9}{143}\rho d_4+\frac{96}{935}\rho d_6\;,
\label{eq:en5d8}
\end{equation}

 \item $D_{6}$

generators: $\{e^{i\pi}\hat{\mathbf{C}}_{2,x},e^{i\pi}\hat{\mathbf{C}}_{6,z}\}$

\begin{equation}
 \zeta_{D6}=\frac{1}{\sqrt{2}}(0,0,1,0,0,0,0,0,-1,0,0)^T
\label{eq:5d6}
\end{equation}
\begin{equation}
 \varepsilon(\zeta_{D10})=\frac{1}{2}\rho c_0+\frac{1}{22}\rho d_0+\frac{1}{1716}\rho d_2+\frac{9}{143}\rho d_4+\frac{2689}{22440}\rho d_6\;,
\label{eq:en5d6}
\end{equation}
 \item $D_{4}$

generators: $\{e^{i\pi}\hat{\mathbf{C}}_{2,x},e^{i\pi}\hat{\mathbf{C}}_{4,z}\}$

\begin{equation}
 \zeta_{D4}=\frac{1}{\sqrt{2}}(0,0,0,1,0,0,0,-1,0,0,0)^T
\label{eq:5d4}
\end{equation}
\begin{equation}
 \varepsilon(\zeta_{D4})=\frac{1}{2}\rho c_0+\frac{1}{22}\rho d_0+\frac{3}{143}\rho d_2+\frac{9}{143}\rho d_4+\frac{96}{935}\rho d_6\;,
\label{eq:en5d4}
\end{equation}
 \item $D_{2}$

generators: $\{e^{i\pi}\hat{\mathbf{C}}_{2,x},e^{i\pi}\hat{\mathbf{C}}_{2,z}\}$

\begin{equation}
 \zeta_{D2}=\frac{1}{\sqrt{2}}(0,0,0,0,1,0,-1,0,0,0,0)^T
\label{eq:5d2}
\end{equation}
\begin{equation}
 \varepsilon(\zeta_{D2})=\frac{1}{2}\rho c_0+\frac{1}{22}\rho d_0+\frac{1}{11}\rho d_2+\frac{9}{143}\rho d_4+\frac{41}{935}\rho d_6\;.
\label{eq:en5d2}
\end{equation}
\end{itemize}

\begin{figure}[h]
 \centering
 \includegraphics[width=6.cm,keepaspectratio=true]{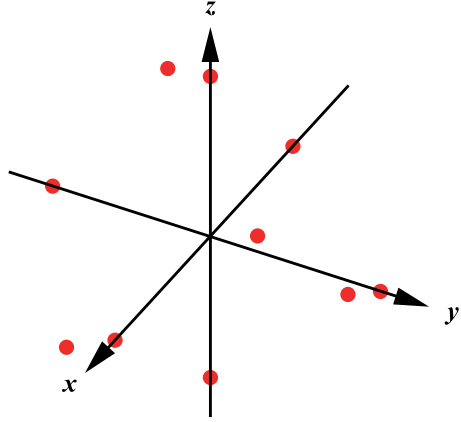}
 \caption{The same as in Fig.~\ref{Fig:5f5} but for the state $\zeta_{T}$, Eq.~(\ref{eq:tetrastan5}).}
 \label{Fig:tf5}
\end{figure}

\begin{figure}[h]
 \centering
 \includegraphics[width=6.cm,keepaspectratio=true]{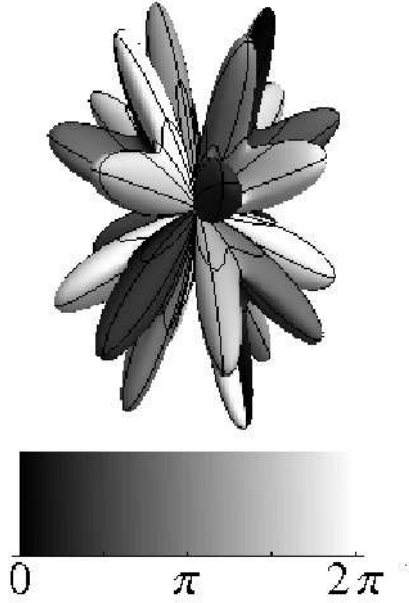}
 \caption{The same as in Fig.~\ref{Fig:5f5sph} but for the state $\zeta_{T}$, Eq.~(\ref{eq:tetrastan5}).}
 \label{Fig:tf5sph}
\end{figure}

Another point group, which should be considered is tetrahedron group $T$ with generators $\hat{C}_{3,x+y+z}$ (rotation about $\vec{x}+\vec{y}+\vec{z}$-axis by angle $2\pi/3$) and $\hat{C}_{2,z}$ (rotation about $\vec{z}$-axis by angle $\pi$). There is only one inert-state whose isotropy group is this group, namely:
\begin{equation}
 \zeta_{T}=\frac{1}{2}(0,1,0,i,0,0,0,-i,0,-1,0)^T\;.
\label{eq:tetrastan5}
\end{equation}
One can  check that
\begin{equation}
 \hat{\mathbf{C}}_{3,x+y+z}\zeta_{T}=e^{i \frac{2\pi}{3}}\zeta_{T}\;,
\label{eq:rownwlasne1}
\end{equation}
\begin{equation}
 \hat{\mathbf{C}}_{2,z}\zeta_{T}=\zeta_{T}\;.
\label{eq:rownwlasne2}
\end{equation}
Thus the generators of the isotropy group are $\{\hat{\mathbf{C}}_{2,z},e^{-i2\pi/3}\hat{\mathbf{C}}_{3,x+y+z}\}$. 
\begin{equation}
 \varepsilon(\zeta_{T})=\frac{1}{2}\rho c_0+\frac{6}{143}\rho d_2+\frac{15}{286}\rho d_4+\frac{168}{935}\rho d_6\;.
\label{eq:en5T}
\end{equation}
The state with tetrahedral symmetry is visualized in Fig.~\ref{Fig:tf5} and Fig.~\ref{Fig:tf5sph}.

There are not any states with the octahedral group as isotropy groups. We know also that maximal number of vertices of a polyhedron which represents states for spin-5 condensates is 10. Since icosahedron has 12 vortices there cannot be any inert-state having the icosahedral group as its isotropy group. Therefore the states presented are all inert-states one can find for spin-5 condensates.


\section{Inert-state for spin-6 condensate with external magnetic field.}
Here we have the same global symmetry group as for the corresponding spin-5 case. The calculations are similar as before and results are the following:
\begin{itemize}
\item $U(1)_{F_z+6\phi}$

generator: $\{e^{6i\phi}e^{-i\phi \hat{f}_z}\}$

\begin{equation}
 \xi_{F6+}=(1,0,0,0,0,0,0,0,0,0,0,0,0)^T
\label{eq:6f6+}
\end{equation}
\begin{equation}
 \varepsilon(\xi_{F6+})=\frac{1}{2}\rho c_0+18\rho c_1+36q-6p\;,
\label{eq:en6f6+}
\end{equation}
\item $U(1)_{F_z+5\phi}$

generator: $\{e^{5i\phi}e^{-i\phi \hat{f}_z}\}$

\begin{equation}
 \xi_{F5+}=(0,1,0,0,0,0,0,0,0,0,0,0,0)^T
\label{eq:6f5+}
\end{equation}
\begin{equation}
 \varepsilon(\xi_{F5+})=\frac{1}{2}\rho c_0+\frac{25}{2}\rho c_1+25q-5p\;,
\label{eq:en6f5+}
\end{equation}

\item $U(1)_{F_z+4\phi}$

generator: $\{e^{4i\phi}e^{-i\phi \hat{f}_z}\}$

\begin{equation}
 \xi_{F4+}=(0,0,1,0,0,0,0,0,0,0,0,0,0)^T
\label{eq:6f4+}
\end{equation}
\begin{equation}
 \varepsilon(\xi_{F4+})=\frac{1}{2}\rho c_0+8\rho c_1+\frac{45}{266}\rho d_8+16q-4p\;,
\label{eq:en6f4+}
\end{equation}
\item $U(1)_{F_z+3\phi}$

generator: $\{e^{3i\phi}e^{-i\phi \hat{f}_z}\}$

\begin{equation}
 \xi_{F3+}=(0,0,0,1,0,0,0,0,0,0,0,0,0)^T
\label{eq:6f3+}
\end{equation}
\begin{equation}
 \varepsilon(\xi_{F3+})=\frac{1}{2}\rho c_0+\frac{9}{2}\rho c_1+\frac{42}{323}\rho d_6+\frac{12}{133}\rho d_8+9q-3p\;,
\label{eq:en6f3+}
\end{equation}
\item $U(1)_{F_z+2\phi}$

generator: $\{e^{2i\phi}e^{-i\phi \hat{f}_z}\}$

\begin{equation}
 \xi_{F2+}=(0,0,0,0,1,0,0,0,0,0,0,0,0)^T
\label{eq:6f2+}
\end{equation}
\begin{equation}
 \varepsilon(\xi_{F2+})=\frac{1}{2}\rho c_0+2\rho c_1+\frac{245}{3431}\rho d_4+\frac{252}{3553}\rho d_6+\frac{18}{247}\rho d_8+4q-2p\;,
\label{eq:en6f2+}
\end{equation}
\item $U(1)_{F_z+\phi}$

generator: $\{e^{i\phi}e^{-i\phi \hat{f}_z}\}$

\begin{equation}
 \xi_{F1+}=(0,0,0,0,0,1,0,0,0,0,0,0,0)^T
\label{eq:6f1+}
\end{equation}
\begin{equation}
 \varepsilon(\xi_{F1+})=\frac{1}{2}\rho c_0+\frac{1}{2}\rho c_1+\frac{21}{286}\rho d_2+\frac{140}{2431}\rho d_4+\frac{210}{3553}\rho d_6+\frac{180}{2717}\rho d_8+q-p\;,
\label{eq:en6f1+}
\end{equation}
\item $U(1)_{F_z}$

generator: $\{e^{-i\phi \hat{f}_z}\}$

\begin{equation}
 \xi_{P}=(0,0,0,0,0,0,1,0,0,0,0,0,0)^T
\label{eq:6p}
\end{equation}
\begin{equation}
 \varepsilon(\xi_{P})=\frac{1}{2}\rho c_0+\frac{1}{26}\rho d_0+\frac{7}{143}\rho d_2+\frac{126}{2431}\rho d_4+\frac{200}{2553}\rho d_6+\frac{175}{2717}\rho d_8\;,
\label{eq:en6p}
\end{equation}
\item $U(1)_{F_z-\phi}$

generator: $\{e^{-i\phi}e^{-i\phi \hat{f}_z}\}$

\begin{equation}
 \xi_{F1-}=(0,0,0,0,0,0,0,1,0,0,0,0,0)^T
\label{eq:6f1-}
\end{equation}
\begin{equation}
 \varepsilon(\xi_{F1-})=\frac{1}{2}\rho c_0+\frac{1}{2}\rho c_1+\frac{21}{286}\rho d_2+\frac{140}{2431}\rho d_4+\frac{210}{3553}\rho d_6+\frac{180}{2717}\rho d_8+q+p\;,
\label{eq:en6f1-}
\end{equation}
\item $U(1)_{F_z-2\phi}$

generator: $\{e^{-2i\phi}e^{-i\phi \hat{f}_z}\}$

\begin{equation}
 \xi_{F2+}=(0,0,0,0,0,0,0,0,1,0,0,0,0)^T
\label{eq:6f2-}
\end{equation}
\begin{equation}
 \varepsilon(\xi_{F2-})=\frac{1}{2}\rho c_0+2\rho c_1+\frac{245}{3431}\rho d_4+\frac{252}{3553}\rho d_6+\frac{18}{247}\rho d_8+4q+2p\;,
\label{eq:en6f2-}
\end{equation}
\item $U(1)_{F_z-3\phi}$

generator: $\{e^{-3i\phi}e^{-i\phi \hat{f}_z}\}$

\begin{equation}
 \xi_{F3+}=(0,0,0,0,0,0,0,0,0,1,0,0,0)^T
\label{eq:6f3-}
\end{equation}
\begin{equation}
 \varepsilon(\xi_{F3-})=\frac{1}{2}\rho c_0+\frac{9}{2}\rho c_1+\frac{42}{323}\rho d_6+\frac{12}{133}\rho d_8+9q+3p\;,
\label{eq:en6f3-}
\end{equation}

\item $U(1)_{F_z-4\phi}$

generator: $\{e^{-4i\phi}e^{-i\phi \hat{f}_z}\}$

\begin{equation}
 \xi_{F4-}=(0,0,0,0,0,0,0,0,0,0,1,0,0)^T
\label{eq:6f4-}
\end{equation}
\begin{equation}
 \varepsilon(\xi_{F4-})=\frac{1}{2}\rho c_0+8\rho c_1+\frac{45}{266}\rho d_8+16q+4p\;,
\label{eq:en6f4-}
\end{equation}

\item $U(1)_{F_z-5\phi}$

generator: $\{e^{-5i\phi}e^{-i\phi \hat{f}_z}\}$

\begin{equation}
 \xi_{F5-}=(0,0,0,0,0,0,0,0,0,0,0,1,0)^T
\label{eq:6f5-}
\end{equation}
\begin{equation}
 \varepsilon(\xi_{F5-})=\frac{1}{2}\rho c_0+\frac{25}{2}\rho c_1+25q+5p\;,
\label{eq:en6f5-}
\end{equation}
\item $U(1)_{F_z-6\phi}$

generator: $\{e^{-6i\phi}e^{-i\phi \hat{f}_z}\}$

\begin{equation}
 \xi_{F6+}=(0,0,0,0,0,0,0,0,0,0,0,0,1)^T
\label{eq:6f6-}
\end{equation}
\begin{equation}
 \varepsilon(\xi_{F6+})=\frac{1}{2}\rho c_0+18\rho c_1+36q+6p\;.
\label{eq:en6f6-}
\end{equation}
\end{itemize}

\section{Inert-states for spin-6 condensates without external magnetic field.}

Similarly as in the corresponding case of spin-5 condensates the inert-states found in the presence of an external magnetic field are also inert-states without the field. Also similarly as previously pairs of these states are actually the same states because in the absence of the field they can be transformed to each other by applying elements of the global symmetry group of the system. These transformations are showed below:
 \begin{equation}
 g(0,\pi,0,0)\xi_{F6+}=\xi_{F6-}
\label{eq:66+=6-z}
\end{equation}
\begin{equation}
 g(0,\pi,0,\pi)\xi_{F5+}=\xi_{F5-}
\label{eq:65+=5-z}
\end{equation}
\begin{equation}
 g(0,\pi,0,0)\xi_{F4+}=\xi_{43-}
\label{eq:64+=4-}
\end{equation}
\begin{equation}
 g(0,\pi,0,\pi)\xi_{F3+}=\xi_{F3-}
\label{eq:63+=3-z}
\end{equation}
\begin{equation}
 g(0,\pi,0,0)\xi_{F2+}=\xi_{F2-}
\label{eq:62+=2-z}
\end{equation}
 \begin{equation}
 g(0,\pi,0,\pi)\xi_{F1+}=\xi_{F1-}\;.
\label{eq:61+=1-z}
\end{equation}

The state $\xi_P$, Eq.~(\ref{eq:6p}), is also an inert-state whose isotropy group is the continuous dihedral group $D_{\infty}$. The other inert-states with the dihedral groups as isotropy groups are the following:
\begin{itemize}
 \item $D_{12}$

generators: $\{e^{i\pi}\hat{\mathbf{C}}_{2,x},e^{i\pi}\hat{\mathbf{C}}_{12,z}\}$

\begin{equation}
 \xi_{D12}=\frac{1}{\sqrt{2}}(1,0,0,0,0,0,0,0,0,0,0,0,-1)^T
\label{eq:6d12}
\end{equation}
\begin{equation}
 \varepsilon(\xi_{D12})=\frac{1}{2}\rho c_0+\frac{1}{26}\rho d_0+\frac{11}{91}\rho d_2+\frac{891}{12376}\rho d_4+\frac{11}{646}\rho d_6+\frac{11}{6916}\rho d_8\;,
\label{eq:en6d12}
\end{equation}

 \item $D_{10}$

generators: $\{e^{i\pi}\hat{\mathbf{C}}_{2,x},e^{i\pi}\hat{\mathbf{C}}_{10,z}\}$

\begin{equation}
 \xi_{D10}=\frac{1}{\sqrt{2}}(0,1,0,0,0,0,0,0,0,0,0,-1,0)^T
\label{eq:6d10}
\end{equation}
\begin{equation}
 \varepsilon(\xi_{D10})=\frac{1}{2}\rho c_0+\frac{1}{26}\rho d_0+\frac{11}{364}\rho d_2+\frac{99}{3094}\rho d_4+\frac{275}{2584}\rho d_6+\frac{275}{6916}\rho d_8\;,
\label{eq:en6d10}
\end{equation}

 \item $D_{8}$

generators: $\{e^{i\pi}\hat{\mathbf{C}}_{2,x},e^{i\pi}\hat{\mathbf{C}}_{8,z}\}$

\begin{equation}
 \xi_{D8}=\frac{1}{\sqrt{2}}(0,0,1,0,0,0,0,0,0,0,-1,0,0)^T
\label{eq:6d8}
\end{equation}
\begin{equation}
 \varepsilon(\xi_{D8})=\frac{1}{2}\rho c_0+\frac{1}{26}\rho d_0+\frac{11}{1001}\rho d_2+\frac{1152}{17017}\rho d_4+\frac{8}{3553}\rho d_6+\frac{3589}{19019}\rho d_8\;,
\label{eq:en6d8}
\end{equation}

 \item $D_{6}$

generators: $\{e^{i\pi}\hat{\mathbf{C}}_{2,x},e^{i\pi}\hat{\mathbf{C}}_{6,z}\}$

\begin{equation}
 \xi_{D6}=\frac{1}{\sqrt{2}}(0,0,0,1,0,0,0,0,0,-1,0,0,0)^T
\label{eq:6d6}
\end{equation}
\begin{equation}
 \varepsilon(\xi_{D6})=\frac{1}{2}\rho c_0+\frac{1}{26}\rho d_0+\frac{25}{4004}\rho d_2+\frac{729}{34034}\rho d_4+\frac{3697}{28424}\rho d_6+\frac{3793}{76076}\rho d_8\;,
\label{eq:en6d6}
\end{equation}
 \item $D_{4}$

generators: $\{e^{i\pi}\hat{\mathbf{C}}_{2,x},e^{i\pi}\hat{\mathbf{C}}_{4,z}\}$

\begin{equation}
 \xi_{D4}=\frac{1}{\sqrt{2}}(0,0,0,0,1,0,0,0,-1,0,0,0,0)^T
\label{eq:6d4}
\end{equation}
\begin{equation}
 \varepsilon(\xi_{D4})=\frac{1}{2}\rho c_0+\frac{1}{26}\rho d_0+\frac{25}{1001}\rho d_2+\frac{537}{10472}\rho d_4+\frac{373}{7106}\rho d_6+\frac{601}{5852}\rho d_8\;,
\label{eq:en6d4}
\end{equation}
 \item $D_{2}$

generators: $\{e^{i\pi}\hat{\mathbf{C}}_{2,x},e^{i\pi}\hat{\mathbf{C}}_{2,z}\}$

\begin{equation}
 \xi_{D2}=\frac{1}{\sqrt{2}}(0,0,0,0,0,1,0,-1,0,0,0,0,0)^T
\label{eq:6d2}
\end{equation}
\begin{equation}
 \varepsilon(\xi_{D2})=\frac{1}{2}\rho c_0+\frac{1}{26}\rho d_0+\frac{99}{1001}\rho d_2+\frac{1002}{17017}\rho d_4+\frac{155}{3553}\rho d_6+\frac{655}{19019}\rho d_8\;.
\label{eq:en6d2}
\end{equation}
\end{itemize}

In the case of the tetrahedron group we have found two candidates for inert-states:
\begin{itemize}
 \item generators: $\{\hat{\mathbf{C}}_{2,z},e^{-i2\pi/3}\hat{\mathbf{C}}_{3,x+y+z}\}$

\begin{equation}
 \xi_{T1}=(\frac{\sqrt{11}}{8},0,\frac{i}{4\sqrt{2}},0,\frac{\sqrt{5}}{8},0,\frac{i\sqrt{7}}{4},0,\frac{\sqrt{5}}{8},0,\frac{i}{4\sqrt{2}},0,\frac{\sqrt{11}}{8})^T,
\label{eq:6t1}
\end{equation}
\begin{equation}
\varepsilon(\xi_{T1})=\frac{1}{2}\rho c_0+\frac{72}{1001}\rho d_2+\frac{15}{17017}\rho d_4+\frac{224}{3553}\rho d_6+\frac{2922}{19019}\rho d_8\;,
\label{eq:en6t1}
\end{equation}
 \item generators: $\{\hat{\mathbf{C}}_{2,z},e^{i2\pi/3}\hat{\mathbf{C}}_{3,x+y+z}\}$

\begin{equation} \xi_{T2}=(\frac{\sqrt{11}}{8},0,-\frac{i}{4\sqrt{2}},0,\frac{\sqrt{5}}{8},0,-\frac{i\sqrt{7}}{4},0,\frac{\sqrt{5}}{8},0,-\frac{i}{4\sqrt{2}},0,\frac{\sqrt{11}}{8})^T,
\label{eq:6t2}
\end{equation}
\begin{equation}
\varepsilon(\xi_{T2})=\frac{1}{2}\rho c_0+\frac{72}{1001}\rho d_2+\frac{15}{17017}\rho d_4+\frac{224}{3553}\rho d_6+\frac{2922}{19019}\rho d_8\;.
\label{eq:en6t2}
\end{equation}
\end{itemize}
The energies of these two states are the same and it turns out that they belong to the same orbit
\begin{equation}
 g(0,0,\pi,\pi)\xi_{T1}=\xi_{T2}\;,
\label{eq:6t1=t2}
\end{equation}
and thus they are actually the same inert-state.

Another group which have to consider is the octahedron group. The octahedron group is group generated by three generators $\hat{C}_{4,z}$ (rotation about $\vec{z}$-axis by angle $\pi/2$), $\hat{C}_{3,x+y+z}$ (rotation about $\vec{x}+\vec{y}+\vec{z}$-axis by angle $2\pi/3$) and finally $\hat{C}_{2,x+y}$ (rotation about $\vec{x}+\vec{y}$-axis by angle $\pi$). For this group we have found one inert-state:
\begin{equation} \xi_{O}=(0,0,-\frac{i\sqrt{7}}{4},0,0,0,\frac{i}{2\sqrt{2}},0,0,0,-\frac{i\sqrt{7}}{4},0,0)^T,
\label{eq:6o1}
\end{equation}
\begin{equation}
\varepsilon(\xi_{O})=\frac{1}{2}\rho c_0+\frac{1}{26}\rho d_0+\frac{1323}{19448}\rho d_4+\frac{4}{3553}\rho d_6+\frac{189}{988}\rho d_8\;.
\label{eq:en6o}
\end{equation}
The $\xi_O$ state is an eigenstate of the $\hat{C}_{4,z}$, $\hat{C}_{3,x+y+z}$ and $\hat{C}_{2,x+y}$ generators corresponding to eigenvalues equal 1. Thus, no phase supplement is necessary because the original generators form the isotropy group of $\xi_O$.

The last group is the icosahedron group. We ought to consider that group because an icosahedron has 12 vertices and maximal number of vertices for illustration of spin-6 states is exactly 12. Icosahedron group's generators are $\hat{C}_{5,z}$ (rotation about $\vec{z}$-axis by angle $2\pi/5$), $\hat{C}_{2,y}$ (rotation about $\vec{y}$-axis by angle $\pi$) and $\hat{C}_{2,(-1+\sqrt{5})x+(2+\sqrt{5})z}$ (rotation about $(-1+\sqrt{5})\vec{x}+(2+\sqrt{5})\vec{z}$-axis by angle $\pi$). For this group we have found one inert-state: 
\begin{equation}
 \xi_{Y}=\frac{1}{5}(0,-\sqrt{7},0,0,0,0,\sqrt{11},0,0,0,0,\sqrt{7},0)^T,
\label{eq:6y}
\end{equation}
\begin{equation}
 \varepsilon(\xi_{Y})=\frac{1}{2} \rho c_0+\frac{1}{26}\rho d_0+\frac{121}{646}\rho d_6\;.
\label{eq:en6y}
\end{equation}
Similarly as in the case of the octahedron group no phase supplement is necessary because the generators $\hat{C}_{5,z}$, $\hat{C}_{2,y}$ and $\hat{C}_{2,(-1+\sqrt{5})x+(2+\sqrt{5})z}$ form the isotropy group of $\xi_Y$.

\section{Conclusion}

In this paper we have presented all inert-states for both spin-5 and spin-6 systems with and without external magnetic field. We have found all these states by considerations of symmetries of energy density functional. Idea of finding inert-states was used in superconductors and Bose-Einstein condensates for lower spins before \cite{he3, supercond1,yip,makela,ueda}. Some inert-states for spin-5 and spin-6 were also presented in \cite{makela}. 

We look for the inert-states employing the Michel's theorem. Yip argues that if one has found inert-states for some group he need not to consider any subgroup of this group because there are no additional inert-states related to the subgroups \cite{yip}. It is obviously truth. We want to pay attention, that any of dihedral groups presented in our paper is subgroup of other. However, M\"{a}kel\"{a} and Suominen in Ref.~\cite{makela} point out that we should not forget about very important global phase changing factors in their generators, which cause that they cannot be subgroups of other. They consider groups $D_4$ and $D_8$. These groups without phase factors obviously obey relation $D_{8}\subset D_{4}$. But in our case it is not true, because:
\[(e^{i\pi}\hat{\mathbf{C}}_{8,z})^2=\hat{\mathbf{C}}_{4,z}\neq e^{i\pi}\hat{\mathbf{C}}_{4,z}.\]
Thus we can easily see that we cannot construct elements of group $D_4$ by using group $D_8$ elements if they are supplemented by phase factors \cite{makela}. Consequently both $D_4$ and $D_8$ has to be considered separately in investigation of inert-states of the systems.
The same argumentation can be repeated for groups $D_2$ and $D_4$. 

The results presented in this paper can be used as a first step for calculating all stationary states of considered systems. Changing Hamiltonians parameters (external magnetic field or scattering lengths) one can also identify quantum phase transitions when ground states of the systems change their symmetries. Dynamical properties of the phase transitions can be analyzed employing the Bogoliubov approach that allows us to investigate stability properties of the stationary states \cite{damski,ueda07,ueda07a}. In order to analyze stability of the inert-states one has to solve the Bogoliubov-de~Gennes equations. These equations, however, can not be solved analytically and numerical calculations are necessary. Because we do not know values of scattering lengths for any spin-5 and spin-6 condensates numerical studies of the stability of the inert-states are postponed for future research.

\section{Acknowledgment}
This work is supported by the
Polish Government within research projects 2009-2012.

\appendix
\section*{Appendix 1}
Values of coefficients of Hamiltonian for spin-5 system.

\[c_0=\frac{1}{19}(25 g_8 -6 g_{10})\]
\[c_1=\frac{1}{19}(g_{10}-g_8)\]
\[d_0=\frac{1}{19}(19 g_0 +36 g_{10} -55 g_8)\]
\[d_2=\frac{1}{19}(19 g_2 +33 g_{10} -52 g_8)\]
\[d_4=\frac{1}{19}(19 g_4 +26 g_{10} -45 g_8)\]
\[d_6=\frac{1}{19}(19 g_6 +15 g_{10} -34 g_8)\;.\]

Values of coefficients of Hamiltonian for spin-6 system.
\[c_0=\frac{1}{23}(36 g_{10}-13 g_{12})\]
\[c_1=\frac{1}{23}(g_{12}-g_{10})\]
\[d_0=\frac{1}{23}(23 g_0 + 55 g_{12}-78 g_{10})\]
\[d_2=\frac{1}{23}(23 g_2 + 52 g_{12}-75 g_{10})\]
\[d_4=\frac{1}{23}(23 g_4 + 45 g_{12}-68 g_{10})\]
\[d_6=\frac{1}{23}(23 g_6 + 34 g_{12}-57 g_{10})\]
\[d_8=\frac{1}{23}(23 g_8 + 19 g_{12}-42 g_{10})\;.\]

\end{document}